\documentclass[twocolumn]{aastex63}

\newcommand\rh{\ensuremath{R_{\mathrm{h}}}}

\usepackage{graphicx}
\definecolor{red}{RGB}{255,0,0}

\usepackage{color,soul}
\usepackage{lineno}

\received{September 30, 2020}
\revised{October 25, 2020}
\accepted{June 30, 2021}
\submitjournal{The Planetary Science Journal}

\shorttitle{Comet Hyperactivity}
\shortauthors{Protopapa et al.}

\begin{document}

\title{Non-detection of water-ice grains in the coma of comet 46P/Wirtanen and implications for hyperactivity}

\correspondingauthor{Silvia Protopapa}
\email{sprotopapa@boulder.swri.edu}

\author[0000-0001-8541-8550]{Silvia Protopapa}
\altaffiliation{Visiting Astronomer at the Infrared Telescope Facility, which is operated by the University of Hawaii under contract 80HQTR19D003 with the National Aeronautics and Space Administration.}
\affiliation{Southwest Research Institute, 1050 Walnut Street, Suite 300, Boulder, CO 80302, USA}

\author[0000-0002-6702-7676]{Michael S. P. Kelley}
\altaffiliation{Visiting Astronomer at the Infrared Telescope Facility, which is operated by the University of Hawaii under contract 80HQTR19D003 with the National Aeronautics and Space Administration.}
\affiliation{Department of Astronomy, University of Maryland, College Park, MD 20742, USA}

\author{Charles E. Woodward}
\altaffiliation{Visiting Astronomer at the Infrared Telescope Facility, which is operated by the University of Hawaii under contract 80HQTR19D003 with the National Aeronautics and Space Administration.}
\affiliation{Minnesota Institute for Astrophysics, University of Minnesota, Minneapolis, MN 55455, USA}

\author[0000-0002-5033-9593]{Bin Yang}
\altaffiliation{Visiting Astronomer at the Infrared Telescope Facility, which is operated by the University of Hawaii under contract 80HQTR19D003 with the National Aeronautics and Space Administration.}
\affiliation{ European Southern Observatory, Alonso de Còrdova 3107, Vitacura, Casilla 19001, Santiago, Chile}


\begin{abstract}
  Hyperactive comets have high water production rates, with inferred sublimation areas of order the surface area of the nucleus. Comets 46P/Wirtanen and 103P/Hartley~2 are two examples of this cometary class. Based on observations of comet Hartley~2 by the Deep Impact spacecraft, hyperactivity appears to be caused by the ejection of water-ice grains and/or water-ice rich chunks of nucleus into the coma. These materials increase the sublimating surface area, and yield high water production rates. The historic close approach of comet Wirtanen to Earth in 2018 afforded an opportunity to test Hartley~2 style hyperactivity in a second Jupiter-family comet.  We present high spatial resolution, near-infrared spectroscopy of the inner coma of Wirtanen.  No evidence for the 1.5- or 2.0-\micron{} water-ice absorption bands is found in six 0.8--2.5~\micron{} spectra taken around perihelion and closest approach to Earth.  In addition, the strong 3.0-\micron{} water-ice absorption band is absent in a 2.0--5.3~\micron{} spectrum taken near perihelion. Using spectroscopic and sublimation lifetime models we set constraints on the physical properties of the ice grains in the coma, assuming they are responsible for the comet's hyperactivity. We rule out pure water-ice grains of any size, given their long lifetime. Instead, the hyperactivity of the nucleus and lack of water-ice absorption features in our spectra can be explained either by icy  grains  on  the  order  of  1~$\mu$m  in  size  with  a small amount of low albedo dust (greater than 0.5\% by volume), or large chunks containing significant amounts of water ice. 
\end{abstract}





\section{Introduction}\label{sec:intro}
The Deep Impact eXtended Investigation (DIXI) flyby of comet 103P/Hartley~2 (hereafter, H2) provided new insight into the role of water ice and its relation to the activity level of comets \citep{AHearn2011}. H2 is a hyperactive comet, i.e., a comet with an extremely high water production rate for the given nucleus size. Cometary activity is commonly expressed in terms of active fraction: the surface area of water ice needed to sustain the activity, generally referred to as ``minimum active area" \citep{Keller1990,Combi2011}, divided by the surface area of the nucleus.  For most comets, the active fraction is typically between a few percent to 30\% (\citealt{AHearn1995}, after correcting for the factor of 2 error noted by \citealt{Knight2021}; see also \citealt{Lis2019}). Hyperactive comets are unusual in that their gas production rates require an active fraction of $\sim$100\% \citep{AHearn1995}. In the specific case of H2, \citet{Combi2011} estimated an active fraction of $\sim$100\% during the 2010 apparition. However, rather than a 100\% active surface, Deep Impact's close up view of H2 on November 2010 revealed heterogeneous surface activity, dominated by a single small active area \citep{AHearn2011}. This area produced bright jets in the inner few kilometers of the coma, enriched in CO$_{2}$ gas and pure water-ice grains on the order of 1~$\mu$m in diameter \citep{Protopapa2014}. These observations of H2 have formed the basis for our current understanding of comet hyperactivity: the ejection of water-ice grains into the coma increases the available surface area for water sublimation, which in turn contributes to the high observed water production rate. However, this remains an hypothesis given that the contribution of the small icy grains to the comet's water production rate was poorly constrained. In addition to the 1-$\mu$m water-ice
grains, a second population of particles was detected: distinct, isolated grains that in visible images surround the nucleus of H2, estimated by their brightness to be
cm-sized or larger \citep{Kelley2013,Kelley2015}. The true composition of these particles is unknown. However, if these would be ice-rich chunks and would behave as mini-comet nuclei with photometric properties similar to the nucleus of H2, they could account for the comet's hyperactivity \citep{Kelley2015}. Therefore, to summarize, sublimating water-ice in the coma, whether in the form of small grains and/or in large chunks of nucleus, could explain the high activity fraction of comet H2.

To test the scenario laid out by Deep Impact's observations at comet H2, and to verify whether the H2 case is an isolated phenomenon or other small hyperactive comets display a water-ice grain halo, we took advantage of the historic close approach between Earth and comet 46P/Wirtanen (2018 December 16, 0.078~au from the Earth and 1.1~au from the Sun, hereafter Wirtanen). Similar to H2, comet Wirtanen: (1) has a sub-km radius \citep[0.56$\pm$0.04 km; ][]{Boehnhardt2002,Lamy2004}; (2) belongs to the Jupiter-family comet dynamical class; and (3) is hyperactive \citep{Lamy1998-Wirtanen,GroussinLamy2003,Lis2019}. Because Wirtanen strikingly resembles comet H2 in dynamical class, size, and activity level, it represents the best target to test the hypothesis of hyperactivity framed by the DIXI observations and therefore to look for evidence of water ice in its coma.

Low resolution reflectance spectroscopy (0.8--5.3~\micron) is ideally suited for detecting water-ice grain halos in cometary comae through the identification of the broad water-ice absorption bands at 1.5, 2.0, 3.0 and 4.5~$\micron$ \citep[e.g.,][]{Fink1982,Warren1984,Mastrapa2008,Mastrapa2009}. The relative strength and shape of these features give us information on the abundance, grain size and purity (i.e., fraction of refractory materials) of the water ice. Direct detection of ice grains in cometary comae via near-infrared spectroscopy has been reported not only by means of in situ observations \citep{Sunshine2007,AHearn2011,Protopapa2014} but also through ground-based facilities \citep{Davies1997,Kawakita2004ApJ,Yang2009,Yang2010,Yang2014,Protopapa2018}. 

Once in the coma, the observability of water-ice grains is controlled by the ice grain sublimation lifetime, which strongly depends on the presence of refractory material (i.e., “impurities”) within the grain, grain size, and heliocentric distance ($\rh$). Theoretical considerations by \citet{Hanner1981} established that dirty ice grains have very short lifetimes at 1~au from the Sun, on the order of a few hundred seconds, depending on their size. Pure ice grains generally have longer lifetimes, varying from hours to hundreds of years for grains of radius of 1~\micron{} and 50~\micron, respectively. This analysis led to the general
understanding that, unless water-ice grains are pure,
their detection is difficult at small heliocentric distances. An example is given by comet C/2013 US$_{10}$ Catalina (hereafter US10): ground-based near infrared spectroscopic measurements revealed clear evidence of water-ice grains in the coma when the comet was at large heliocentric distances ($\rh\geq3.9$~au), while a featureless spectrum was recorded at $\rh\lesssim2.5$~au \citep{Protopapa2018}. The changing ice fraction in the coma has been interpreted by \citet{Protopapa2018} in terms of the limited lifetime of the water-ice grains
with respect to the field-of-view when closer to the Sun. Under the assumption that the ice
was continuously ejected from the nucleus, the non-detection of water-ice absorption bands close to perihelion enabled them to further constrain the physical properties of the grains and rule out the case of pure icy grains. Micrometer-sized grains of water ice mixed with a small fraction of low-albedo dust, ~0.5\% by volume, were found to be consistent with all US10 spectroscopic measurements. Additionally, based on the presence of a CO$_{2}$ coma, the authors put forth the idea of US10 being a possible hyperactive comet, although the nucleus size is still unknown. 

The close approach of comet Wirtanen to Earth enabled exceptional spatial scale observations (1\arcsec{}$\geq$60~km), which are critical for detecting water ice at 1~au from the Sun, given  the limited water-ice lifetime at such heliocentric distances. Specifically, 1-$\mu$m pure water-ice grains similar to those observed by Deep Impact at comet H2 should be detectable from the ground in the coma of Wirtanen during the comet's closest approach to Earth (see Section \ref{subsec:Water-ice grain sublimation lifetime}). Nevertheless, as noted by \citet{AHearn1981}, even a non detection of solid state water in spectroscopic measurements does not rule out the presence of ice. Instead, when combined with ice sublimation models, it enables us to set constraints on the physical properties of the water ice.

Here, we present NASA Infrared Telescope Facility (IRTF) SpeX observations of the hyperactive comet Wirtanen during its closest approach to Earth (Section \ref{sec:Observations}) acquired with the objective of addressing  the question: can the present model for hyperactivity based on comet H2 be generalized to other hyperactive comets? A series of near-infrared spectroscopic observations of comet Wirtanen were collected to monitor and assess the composition of ejected grains as insolation varies on the nucleus, due to changes in heliocentric distance, and as the comet undergoes vigorous coma activity. The spectral results and the sensitivity to detect water ice are discussed in Section \ref{subsec:Qualitative analysis of the spectral results}, while the contribution of the nucleus to the observed flux density is presented in Section \ref{subsec:nucleus}. The spectral modeling used to interpret the data and the constraints derived on the dust properties are outlined in Sections \ref{subsec:Modeling} and \ref{subsec:Ice free model: Properties of the dust coma}, respectively. Sections \ref{subsec:Upper limit for the water-ice abundance} and \ref{subsec:Water-ice grain sublimation lifetime} provide constraints on the relative abundance and physical properties (i.e., purity and grain size) of the water-ice grains in the coma based on spectroscopic and sublimation lifetime models, respectively. Specifically, we test for the presence of water-ice grains with physical properties identical to those detected in the coma of H2 and US10, a confirmed and hypothetical hyperactive comet, respectively. The lack of variation in the color of the dust across our observational
data set is discussed in Section \ref{subsec:Coma reddening}. A summary and discussion follows in Section \ref{sec:Discussion and conclusions}.

\begin{figure*}[htp]
  \includegraphics[width=\linewidth]{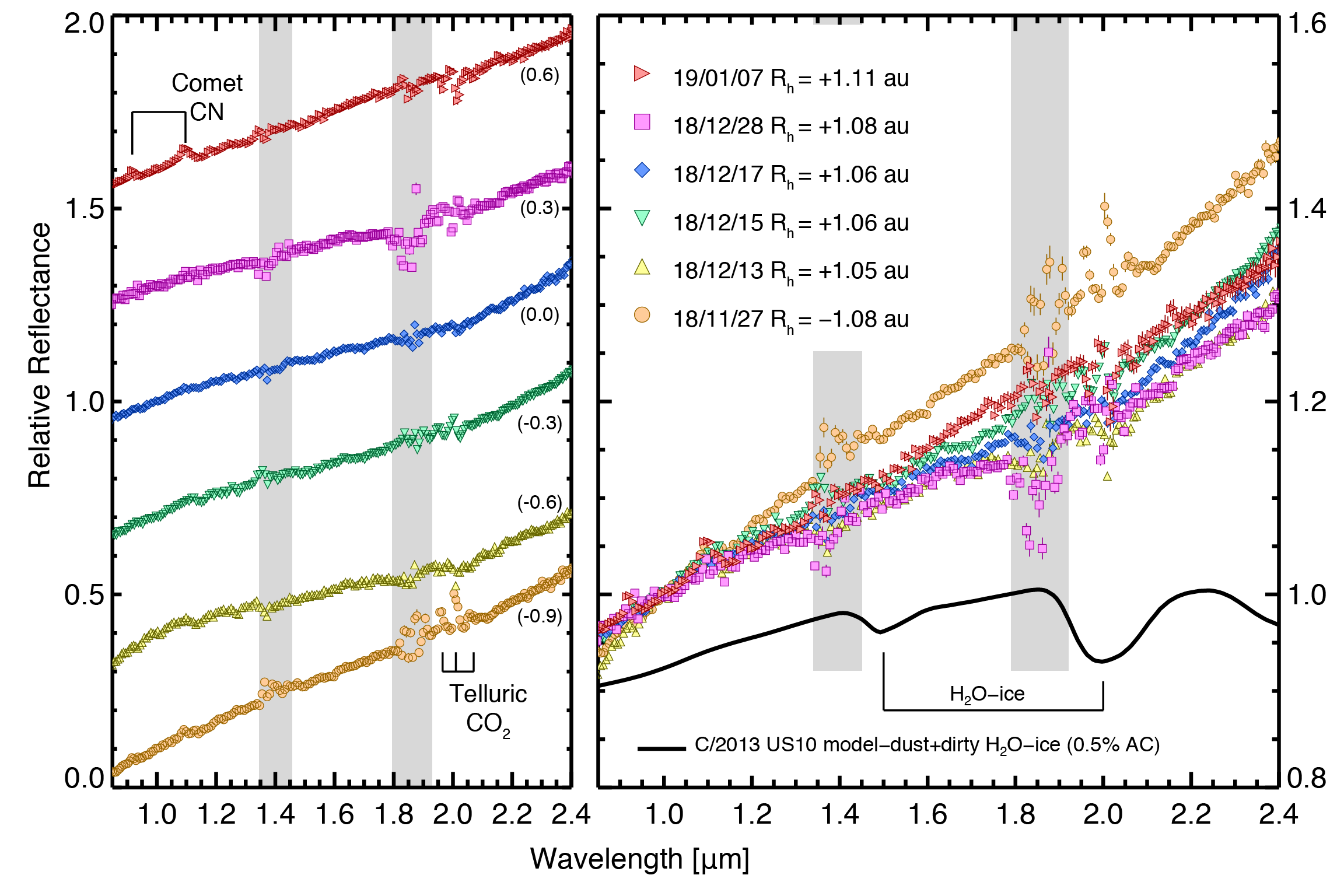}
  \caption{IRTF SpeX prism mode observations of Wirtanen acquired at $|\rh|$ between 1.05--1.11~au, binned to a spectral resolution of $\sim$300 at 2.0~\micron. Right panel: Spectra are normalized between 0.98 and 1.02~\micron{} to unity. Overplotted is the best fit model (solid line) of the C/2013 US$_{10}$ observations acquired at $|\rh|\geq$3.9~au consistent with the presence of water-ice grains of uniform size, with a diameter of 1.2~\micron, and a coma ice-to-dust ratio of 0.18 \citep{Protopapa2018}. Left panel: Spectra are normalized to unity between 0.98 and 1.02~\micron{} and shifted along the y-axis by values shown in parentheses. The gray shaded rectangles are regions of strong telluric absorption. The spectra show no water-ice absorption features. The data used to create this figure are available. \label{fig: prism_observations}}
\end{figure*}

\begin{deluxetable*}{ccccccccccc}
  \tablecaption{Observational Parameters for 46P/Wirtanen.\label{tab_obs}}
  \setlength{\tabcolsep}{3pt} 
  \tablecolumns{14}
  \tablewidth{0pt}
  \tablehead{
    \colhead{UT date} &
    \colhead{Mode} &
    \colhead{Slit\tablenotemark{a}} &
    \colhead{AM$_{target}$\tablenotemark{b}} &
    \colhead{SA (spectral type)\tablenotemark{c}} &
    \colhead{AM$_{SA}$\tablenotemark{d}} &
    \colhead{\rh{}\tablenotemark{e}} &
    \colhead{$\Delta$\tablenotemark{f}} &
    \colhead{T-mag\tablenotemark{g}} &
    \colhead{S\tablenotemark{h}} &
    \colhead{STO\tablenotemark{i}}\\
    \colhead{(YYYY-MM-DD HR:MN)} &
    \colhead{} &
    \colhead{(arcsec)} &
    \colhead{} &
    \colhead{} &
    \colhead{} &
    \colhead{au} &
    \colhead{au} &
    \colhead{} &
    \colhead{\%/(100 nm)} &
    \colhead{deg}\\\\
  }
  \startdata
  \hline
  2018-11-27 06:55 & Prism	 & 0.8$\times$15 	& 1.51$\pm$0.03 & HD 17883 (G2V C) & 1.64$\pm$0.01 & -1.076 & 0.14 & 12.3 & --- & 46.5\\
  2018-12-13 06:55 & Prism	 & 0.8$\times$60 	& 1.10$\pm$0.03 & HD 25680 (G5V C) & 1.04$\pm$0.01 & +1.055 & 0.08 & 11.1 & 1.9 & 26.4\\
  2018-12-13 08:36 & LXD$\_$l  	& 0.8$\times$15 & 1.01$\pm$0.01 & HD 21379 (A0V C) & 1.03$\pm$0.01& +1.055 & 0.08 & 11.1 & & 26.3\\
  2018-12-15 07:08& Prism 	& 0.8$\times$60 	& 1.07$\pm$0.03 & HD 30246 (G5 D)    & 1.11$\pm$0.01 & +1.056 & 0.08 & 11.1 & 2.3 &22.2\\
  2018-12-17 06:20& Prism 	& 0.8$\times$60 	& 1.21$\pm$0.01 & HD 30246 (G5 D)    & 1.16$\pm$0.01 & +1.057 & 0.08 & 11.1 & 2.1 & 19.2\\
  2018-12-28 07:03& Prism 	& 0.8$\times$60 	& 1.55$\pm$0.02 & HD 30246 (G5 D)    & 1.51$\pm$0.01 & +1.076 & 0.10 & 11.8 & 1.9 & 27.7\\
  2019-01-07 12:51& Prism 	& 0.8	$\times$60 	& 1.36$\pm$0.01 & HD 292561 (F8 E)   & 1.31$\pm$0.01 &  +1.110& 0.16 & 12.7 & 2.5 & 33.2\\
  \hline
  \enddata
  \tablenotemark{a}{Slit width $\times$ slit length.}
  \tablenotemark{b}{Average airmass of the comet observations.}
  \tablenotemark{c}{Star used for calibration purposes, its spectral type and precision from SIMBAD.}
  \tablenotemark{d}{Average airmass of the solar type star observations.}
  \tablenotemark{e}{The Sun-to-target distance ($<$0~au indicates pre-perihelion epochs).}
  \tablenotemark{f}{The target-to-observer distance.}
  \tablenotemark{g}{Comet's approximate apparent visual total magnitude as reported by Horizons.}
  \tablenotemark{h}{Comet spectral slope (at 1~\micron). The uncertainty is 0.3~\%/100~nm (see text for details).}
  \tablenotemark{i}{The Sun-Target-Observer angle as reported by Horizons, which differs from the true phase angle at the few arcsecond level}.
\end{deluxetable*}
\section{Observations}\label{sec:Observations}
Observations of Wirtanen were obtained using the IRTF/SpeX \citep{Rayner1998,Rayner2003} high-throughput low-resolution prism mode covering the wavelength range 0.7--2.52~\micron{} (Fig.~\ref{fig: prism_observations}, Table \ref{tab_obs}), under program IDs 2018B006, 2018B088, 2018B132. All observations were acquired with the 0\farcs8$\times$15\arcsec{} or 0\farcs8$\times$60\arcsec{} slit configurations, resulting in a spectral resolving power ($R=\lambda/\Delta\lambda$) of $\sim$112 at 2.0~\micron. We reduced the data using IDL codes, based on Spextool \citep{Cushing2004}. The latter does not reduce observations taken with the 60\arcsec~long slit, which was used extensively in our observing campaign to optimize the background subtraction in the standard infrared method of moving the object between two positions A and B on the slit. Given the high activity of the comet and extended coma, the 60\arcsec~long slit prevented the overlapping of the cometary coma in the A and B pair images. Spectra were extracted, then wavelength calibrated and combined using a robust mean with a 2.5$\sigma$-clipping threshold. The uncertainty is given by the standard deviation on the good pixels. Calibrated spectra were binned by 3 pixels to yield a spectral resolution ($\lambda/\Delta\lambda$, where $\Delta\lambda$ represents the spectral sampling) of $\sim$300 at 2.0~\micron{} (Fig.~\ref{fig: prism_observations}). The spectral extraction aperture was set to the limit beyond which the object’s signal does not improve and only the noise increases. This means that the choice of aperture size is a compromise between statistical noise and photometric accuracy. The spectral extraction radius varied between 1\arcsec~on 2018 November 27 and 7\arcsec~close to perihelion.

\begin{figure}[htp]
  \includegraphics[width=\linewidth]{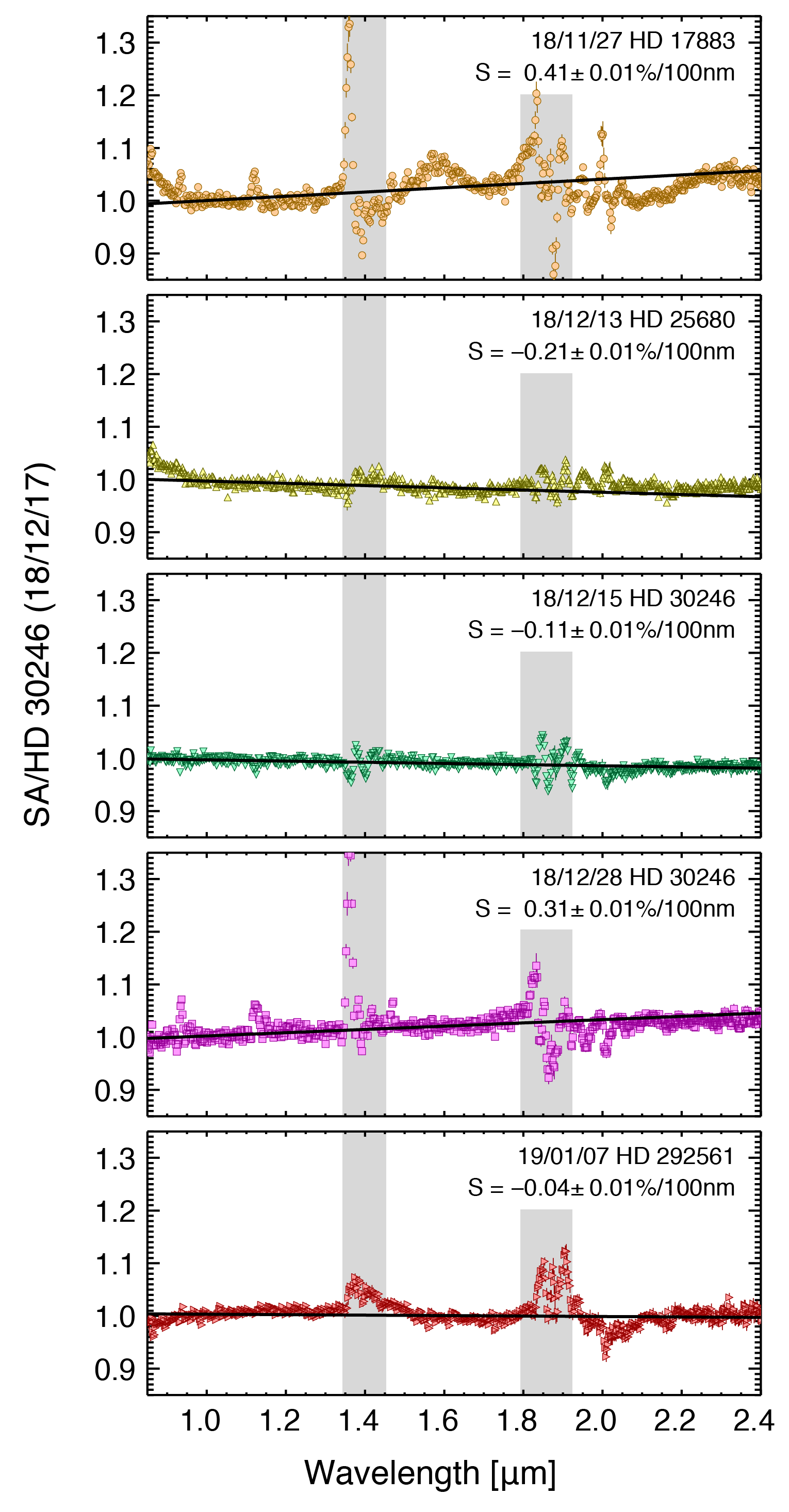}
  \caption{Spectral variations among the solar type reference stars observed during the Wirtanen campaign. The ratio between each solar type star (SA) and a reference one, arbitrarily chosen, HD 30246 (G5 D) observed on 2018 December 17, is displayed, normalized to unity between 0.98 and 1.02~$\mu$m. Overlaid on each ratio spectrum is the linear model (solid black line). The spectral slope resulting from the linear fit, expressed in \%/100~nm, is reported in each panel. \label{fig:prism_observations_SA}}
\end{figure}

To remove telluric absorptions, instrumental effects, and the solar spectrum, all Wirtanen observations were interspersed between observations of solar type stars. The comet-to-solar-type-star ratio produced a spectrum proportional to the comet reflectance. We notice that the correction of telluric features in our comet spectra (Fig.~\ref{fig: prism_observations}) is sometimes imperfect not only near 1.4 and 1.9~\micron, where telluric H$_{2}$O vapor absorption occurs, but also near 2.0~\micron, where lies strong and narrow telluric CO$_{2}$ absorptions (marked in the figure), which make sky transparency variable in time. It was not possible to always observe the same star, and stars with spectral type G2, G5, and F8 were adopted for calibration purposes (Table \ref{tab_obs}). To investigate the spectral variations introduced by the different calibration stars, we have compared the ratios of each solar type star to a common reference, HD 30246 \citep[G5;][]{Russel2007} observed on 2018 December 17 (Fig.~\ref{fig:prism_observations_SA}). The ratio is performed after airmass-correcting each solar type spectrum using telluric absorption models from the Planetary Spectrum Generator \citep{Villanueva2018}. The spectral gradient of each solar type star with respect to HD 30246 is reported in Fig.~\ref{fig:prism_observations_SA} and estimated through a linear fit of the ratio spectrum, normalized to unity at 1~$\mu$m,  in the wavelength ranges between 0.95--1.05, 1.165--1.32, 1.45--1.75, and 2.1--2.2~$\mu$m to avoid spectral regions contaminated by strong telluric absorptions. Most calibration stars present a similar spectral behavior to that of HD 30246, apart from a slight spectral slope on the order of $\sim$0.2\%/100~nm (at 1~$\mu$m). The standard deviation among these spectral gradients equals 0.3\%/100~nm and is adopted as the uncertainty on the comet spectral slopes reported in Table \ref{tab_obs}. The latter are also estimated through a linear fit of the comet spectrum, normalized to unity at 1~$\mu$m,  in the wavelength ranges between 0.95--1.05, 1.165--1.32, 1.45--1.75, and 2.1--2.2~$\mu$m to avoid spectral regions contaminated by strong telluric absorptions. The calibration star HD 17883 observed on 2018 November 27 is the only one that does not display a linear featureless spectrum after normalization to HD 30246, but instead presents broad spectral features at 1.6~$\mu$m and 2.1~$\mu$m, as an example. However, because the corresponding comet spectrum does not display those features (Fig.~\ref{fig: prism_observations}) and a different slit configuration was adopted when observing HD 17883 and HD 30246 on 2018 November 27 and  2018 December 17, respectively (Table \ref{tab_obs}), we attribute these features to instrumental effects. We conclude that all calibration stars in our analysis are good solar type stars and the deviation of their spectral behavior from that of the Sun is accounted for in the comet spectral gradient uncertainty.

The 2018 November 27 data are the only prism-mode spectra in our observing campaign acquired with the 15\arcsec~slit length, and are the only data for which the subtraction between frames taken in the A and B slit positions resulted in a positive and negative image of the coma overlapping each other. This overlap prevents an accurate sky background subtraction in the data reduction, which might affect the spectral slope. Therefore, while the resulting spectrum is reliable to search for the presence of water-ice features, we do not report its spectral gradient in Table \ref{tab_obs}.

The comparison between each comet spectrum (Fig.~\ref{fig: prism_observations}) and that of the corresponding solar type reference star (Fig.~\ref{fig:prism_observations_SA}) enables us to infer whether absorption or emission features observed in the comet reflectance spectrum are to be attributed to the comet or to the calibration star instead. The comet spectrum recorded on 2019 January 07 (red filled rightfacing triangles in Fig.~\ref{fig: prism_observations}) presents two clear emission features which originate from CN. The
strongest of these bands is the 0-0 transition at 1.097~$\mu$m, followed by the 1-0 transition at 0.917~$\mu$m \citep{Johnson1983}. Because these features do not appear in the solar type spectrum (Fig.~\ref{fig:prism_observations_SA}, bottom panel), we attribute them to the comet.

\begin{figure*}[htp]
  \includegraphics[width=\linewidth]{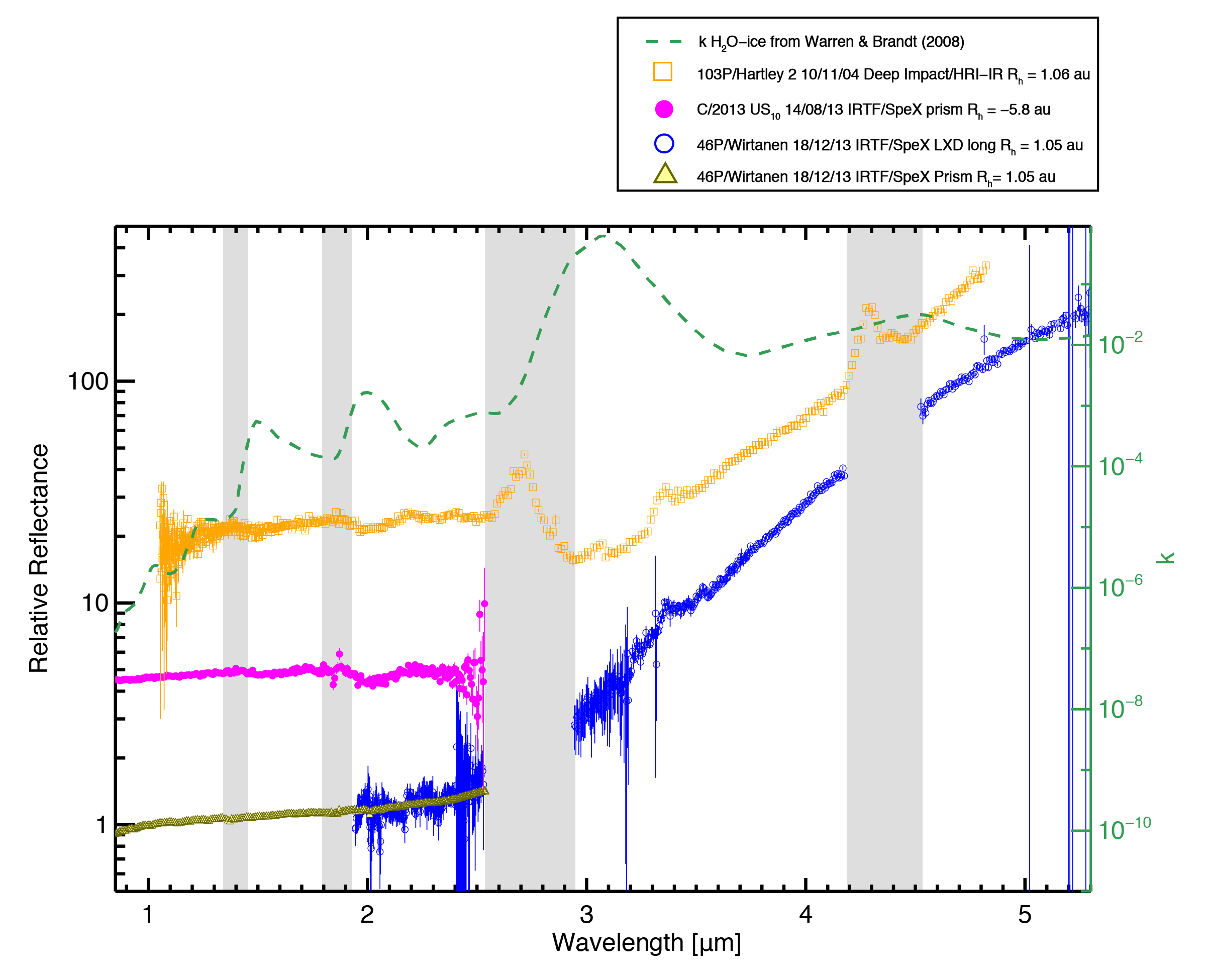}
  \caption{Comet Wirtanen reflectance spectra acquired a few hours after perihelion over the wavelength range 0.8–5.3~$\mu$m. The LXD\_long spectrum (blue open circles) has been normalized between 2.1 and 2.3~$\mu$m to match the prism observations (yellow filled triangles). The spectrum is featureless but for molecular/organic emission at $\sim$3.3~\micron, and residual telluric features. The imaginary part of the refractive index $\textrm{k}$ of crystalline water ice by \citet[][a peak in $\textrm{k}$ corresponds to an absorption band]{Warren2008}, is shown for comparison (green dashed line). No water-ice absorption features are observed in the spectrum of comet Wirtanen throughout the full wavelength range. The Deep Impact/HRI-IR spectrum of 103P/Hartley~2 \citep{Protopapa2014} and the IRTF/SpeX spectrum of C/2013 US10 \citep{Protopapa2018}, which clearly exhibit the characteristic water-ice spectral features, are shown also for comparison. The spectra of 103P/Hartley~2 and C/2013 US10 have been normalized between 2.15 and 2.25~$\mu$m to match the prism observations of 46P/Wirtanen and shifted along the y-axis for clarity.
  \label{fig:comparison46PH2US10}}
\end{figure*}

On 2018 December 13, immediately post perihelion (perihelion date as reported by the MPC occurred on 2018 December 12 at 22:33:34 UT, JD 2458465.43998, 1.055~au from the Sun), observations were also carried out with the long wavelength cross-dispersed (LXD\_long) mode covering the wavelength range 1.98--5.3~\micron{} ($R\sim840$ at 4.5~$\mu$m for the 0\farcs8$\times$15\arcsec{}~slit).  The comet was centered in the slit and the sky background measured by nodding off source.  The LXD\_long data (Fig.~\ref{fig:comparison46PH2US10}) were extracted with a 2\arcsec{} radius aperture, calibrated with a nearby A0V star, following \citet{Vacca2003}, and normalized by the solar spectrum assembled by \citet{Willmer2018}. The latter is a composite spectrum of the Sun, combining empirical spectra assembled by \citet{Haberreiter2017} up to 2~$\mu$m and a model by \citet{Fontenla2011} between 2.0~$\mu$m and 100~$\mu$m. The calibrated spectrum was binned by 8 pixels to yield a spectral resolution equal to the resolving power of $\sim$840 at 4.5~$\mu$m.

\section{Analysis and Results}\label{test}
\subsection{Qualitative analysis of the spectral results}\label{subsec:Qualitative analysis of the spectral results}
Water ice has absorption bands in the near-infrared at 1.5, 2.0, 3.0, and 4.5~\micron{}. The strongest of these absorption features is the 3.0-\micron{} band.  Based on a visual inspection of the prism (0.8--2.4~\micron{}) spectra in Fig.~\ref{fig: prism_observations}, we find no evidence for the 1.5- or 2.0-\micron{} bands.  To emphasize this point, we show, for comparison purposes only, the model spectrum of the icy coma of comet US10 from \citet{Protopapa2018}. The US10 model consists of a heterogeneous coma of dust grains and water-ice grains, where the ice grains are an intimate mixture of amorphous carbon and water ice (0.5\% carbon by volume) and have a best-fit ice grain diameter of 1.2~\micron{}. However, the difference in the 1.5- and 2.0-\micron{} band shapes between 0.5\% dirty ice and pure ice is subtle (see Fig.~3 of \citealt{Protopapa2018}). The LXD\_long spectrum of comet Wirtanen (Fig.~\ref{fig:comparison46PH2US10}) also lacks the 2.0-, 3.0-, and 4.5-\micron{} water-ice absorption bands. However, thermal emission from the nucleus is substantial in the LXD\_long spectrum (see Section~\ref{subsec:nucleus}); it dilutes the spectral signature of water ice at 3.0~\micron, and likely overwhelms any feature at 4.5~\micron{} (see the imaginary part of the refractive index of water ice in Figure~\ref{fig:comparison46PH2US10}, green dashed line). 

The non-detection of water-ice spectral features in the IRTF/SpeX measurements of Wirtanen can not be attributed to the sensitivity of the instrument. These features are readily available with IRTF/SpeX and they were already detected in the coma of several comets observed with the same instrument and spectroscopic modes adopted for the Wirtanen spectroscopic measurements presented here \citep{Yang2009,Yang2013,Yang2014,Protopapa2018}.

The detection of cometary water-ice features depends on 1) spectral resolution and 2) signal-to-noise ratio (SNR) of the observations. Low spectral resolutions, $R$ = 100--200, with SNR of $\sim$30 per bin are sufficient to detect water-ice absorption bands of a few percent by depth in the near-infrared wavelength range. As an example, the Deep Impact spectrum of comet H2 has a spectral resolution of 240 at 2.0~$\mu$m with SNR around 40 in the 1.8--2.5~$\mu$m range, and a 2.0-$\mu$m water-ice band depth of $\sim$10\% \citep[Figure \ref{fig:comparison46PH2US10}, orange squares,][]{Protopapa2014}. IRTF/SpeX observations of US10 acquired in prism mode on 2014 Aug 13 have R$\sim$300 at 2.0~$\mu$m and a SNR of $\sim$30 in the range 1.8--2.5~$\mu$m \citep[Figure~\ref{fig:comparison46PH2US10}, magenta filled dots,][]{Protopapa2018}. Therefore, the IRTF/SpeX and Deep Impact/HRI-IR observations of US10 and H2, respectively, are comparable in terms of sensitivity. Observations of US10 display a water-ice band depth at 2.0~$\mu$m of $\sim$10\%, similar to that observed in the spectrum of comet H2. The spectroscopic data of comet Wirtanen presented here all have the sensitivity to detect US10-like water-ice absorption features. They have the same resolving power as the US10 measurements and an SNR of 320 in the range 1.8--2.5~$\mu$m, which is a factor of ten larger than the US10 spectral data.

With spectral sensitivity ruled out, the non-detection of water-ice absorption bands has to be attributed either to the lack of any solid ice in the coma, or to the presence of water-ice grains with physical properties (grain size and purity) that limit their lifetime with respect to the field- of-view at the time of the observations. The first case is explored in Section \ref{subsec:Ice free model: Properties of the dust coma}, while the second is addressed in Sections \ref{subsec:Upper limit for the water-ice abundance} and \ref{subsec:Water-ice grain sublimation lifetime}.

To investigate the properties of both refractories and ice in the coma of Wirtanen,  we combined the LXD\_long and prism data collected on the same date, only a few hours apart, yielding a nearly complete spectrum from 0.8 to 5.3~\micron{} (Figs.~\ref{fig:comparison46PH2US10} and \ref{fig:prism_lxd_observations}). The LXD\_long spectrum, normalized to match the prism measurements between 2.1 and 2.3~\micron, is in near agreement with the short wavelength data between 2.0 and 2.4~\micron.
\vspace{1cm}
\subsection{Estimate of the nucleus contribution}\label{subsec:nucleus}
The close approach to the Earth made for a rare set of circumstances, and the possibility that the nucleus makes a significant contribution to the observed flux density. We modeled the reflected light and thermal emission from the nucleus with the $H,G$ model and the Near-Earth Asteroid Thermal Model (NEATM), respectively \citep{Bowell1989, Harris1998}. The $H,G$ model parameters and assumed values are effective radius, 555~m, geometric albedo, 4\% at 0.55~\micron{} (used to compute the absolute magnitude $H$), and phase function parameter $G$ \citep[0.15 assumed; see][]{Bowell1989}.  In addition, we redden the nucleus with the same spectral gradient as measured from the near-infrared coma.  The NEATM parameters include all of the $H,G$ parameters, mean infrared emissivity, 0.95, and infrared beaming parameter, 1.03.  The emissivity and beaming parameter values are based on the results for cometary nuclei by \citet{Fernandez2013}.  With these parameters, we compute a model spectrum of the nucleus and compare it to the SpeX LXD data.  We find that the nucleus could account for 15, 35, 61, and 68\% of the spectral flux at 2, 3, 4, and 5~\micron, respectively.  The thermal emission from the comet appears to be dominated by the nucleus.  In contrast, the scattered component is dominated by the coma.

In principle, one could subtract the model nucleus spectrum from the LXD spectrum and analyze the residuals as a spectrum of the coma. However, we made several assumptions in order to arrive at our nucleus estimate, and there is enough uncertainty in the parameters that deriving a useful thermal emission spectrum of the nucleus is not possible. The most uncertain parameter is the size of the nucleus, derived under the most unlikely assumption of a spherical nucleus. \citet{Lamy1998-Wirtanen} and \citet{Boehnhardt2002} found peak-to-peak amplitudes of 0.2 and 0.4~mag, respectively.  Given that the shape and rotational state of the nucleus is unknown at the time of our observations, we immediately have an uncertainty of $\pm$20\% on our nucleus estimate at all wavelengths. Furthermore, the shape of the LXD spectrum relative to the nominal nucleus spectrum is unexpected. Cometary dust comae should have cooler spectral temperatures than cometary nuclei, due to the fact that small grains can re-radiate absorbed sunlight isotropically, but nuclei tend to re-radiate that energy from the sun-lit hemisphere only.  In contrast to the expectations, our nucleus model suggests an increasing nucleus fraction in the LXD spectrum from 3 to 5~\micron{}, which indicates the nucleus has a cooler spectrum. The spectral temperature of the NEATM is controlled by the beaming parameter.  We based our assumed value on the results of the photometric survey by \citet{Fernandez2013}, who find a population averaged beaming parameter of 1.03$\pm$0.11 based on 16- and 22-\micron{} photometry.  By decreasing the beaming parameter, we could produce a warmer spectral temperature which would lead to a higher nucleus contribution.  A 2-$\sigma$ variation of the \citet{Fernandez2013} beaming parameter results in a spectrum that accounts up to 110\% of the LXD spectrum at $>3.5$~\micron.  However, the contribution at 2~\micron, which is dominated by reflected light, is still 15\%.

Given the likely high contribution of the nucleus, we cannot separately analyze the thermal emission of the dust.  We instead elect to characterize the thermal emission with an empirical model, e.g., the Planck function, which allows us to avoid a physical interpretation of the thermal emission, but still is a good description of the data (as we show below).  For the scattered light, we account for the nucleus by assuming it contributes 15\% of the flux, i.e., the remaining 85\% is the coma.
\subsection{Modeling}\label{subsec:Modeling}
The reflectance spectrum of the coma in the wavelength range covered by the prism and LXD\_long data combined is modeled as the sum of the scattered and thermal components of the coma. In this section we describe the general philosophy of the modeling approach.

The thermal component of the reflectance spectrum is approximated by
\begin{equation}
  R_{thermal} = \frac{\pi\:R_{h}^{2}\:B(T_{c}, \lambda)f}{F_{\sun}},
\end{equation}
where $B(T_{c}, \lambda)$ is the Planck function, which depends on the color temperature $T_{c}$, $F_{\sun}$ is the solar flux at 1 au, $R_{h}$ is the comet heliocentric distance in au, and $f$ is the filling factor of the grains. The latter is obtained by normalizing the thermal component of the modeling to the data at 3.9--4.1~\micron{} after subtracting the scattered component. 

The scattered component of the spectrum is modeled as an areal mixture of two components, one of polydisperse porous dust particles, or equivalently dust aggregates, and one of polydisperse water-ice aggregates, to which we refer to as type-A and type-B aggregates, respectively.

We consider the type-A and type-B components separated from one another, such that

\begin{equation}\label{eq:areal_mixture}
  R_{scattered}=\sum_{j}F_{j}r_{j}(\lambda)=F_{A} r_{A}+(1-F_{A})r_{B}
\end{equation}

\noindent where $F_{j}$ is the fraction of the area occupied by the $j$th component with
$\sum_{j}F_{j}=1$, and the subscripts A and B denote the type-A aggregates and type-B grains, respectively.

The reflectance of the $j$th-type of aggregates is approximated by diffusive reflectance and is given by
\begin{equation}
r_{j}(\lambda)=\frac{1-\sqrt{(1-\textrm{w}_j(\lambda))}}{1+\sqrt{(1-\textrm{w}_{j}(\lambda))}},
\end{equation}
where $\textrm{w}_{j}$ is the single scattering albedo of the $j$th type of aggregates. This is 
\begin{equation}
  \textrm{w}_{j}(\lambda) = \frac{\int_{a_{1}}^{a_{2}}\,a^2\:Q_{S}(a,m(\lambda))\:n(a) da}{\int_{a_{1}}^{a_{2}}\,a^2\:Q_{E}(a,m(\lambda))\:n(a) da}
\end{equation}
where $a_{1}$ and $a_{2}$ are the smallest and largest particle radii in the size distribution. $Q_{S}(a,m)$ and $Q_{E}(a,m)$ are the scattering and extinction efficiencies of the aggregates as a function of radius $a$ and complex refractive index $m$, respectively. These are computed by means of Mie scattering theory, so that we can model grains with any size parameter ($\textrm{X}=2\pi a$/$\lambda$), even those outside the geometric optics regime ($\textrm{X}<1$). The real and imaginary parts of the refractive index as a function of $\lambda$ are estimated from effective medium theory \citep[Bruggeman mixing formula,][]{Bohren1983}. We adopt a fractal model for the internal structure (i.e., porosity) of the grains.  Thus, the fractional volume filled by material in a porous grain is given by
\begin{equation}
\delta = (a/a_{0})^{D-3}
\end{equation}
where $a_{0}$ is the radius of the basic structural unit of the grain, which we set equal to 0.1 $\mu$m \citep[e.g.,][]{Harker2002}, and $D$ is the fractal porosity, which ranges between 2.5 and 3 \citep{Lisse1998}. Solid spheres are represented by $D=3$, while for porous spheres $D<$3 \citep{Harker2002}.
The differential particle size distribution $n(a)da$ is the number of particles per unit volume with radius between $a$ and $a+da$, following the formula
\begin{equation}
  n(a) \propto{} a^{-\alpha}.
\end{equation}
\begin{figure*}[htp]
  \plotone{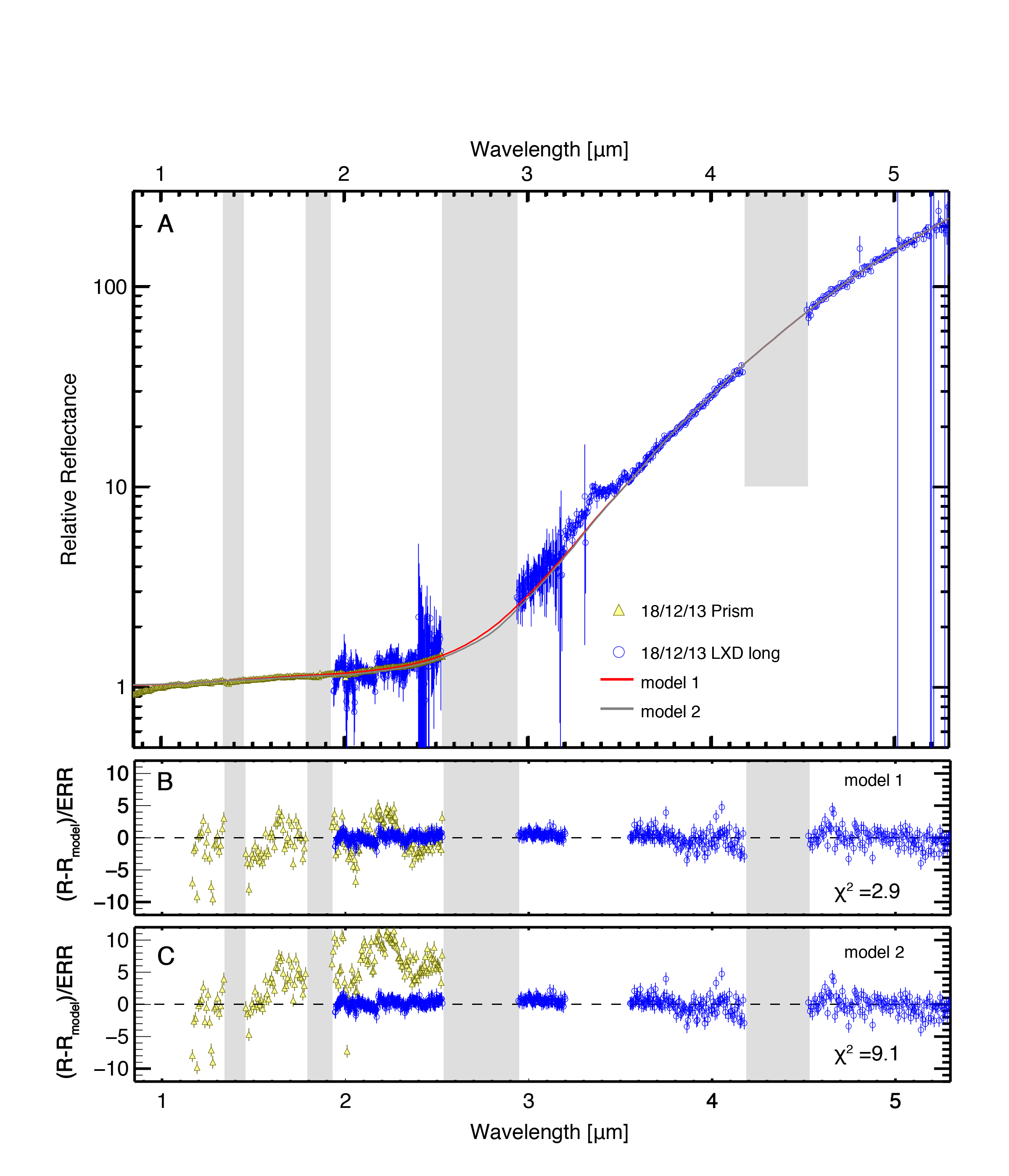}
  \caption{(A) Comet Wirtanen reflectance spectra over the wavelength range 0.8–5.3~$\mu$m, identical to those shown in Figure \ref{fig:comparison46PH2US10}. The ice-free (\textit{model 1}) and ice-rich (\textit{model 2}) models, with an upper limit of 2.2\% for the water-ice abundance, are shown in red and gray solid lines, respectively. Both models do not fit the emission feature as expected. (B and C) Residuals of each fit, with the resulting $\chi^{2}_{\nu}$ indicated in each case. \label{fig:prism_lxd_observations}}
\end{figure*}

\begin{figure*}[htp]
  \includegraphics[width=\linewidth]{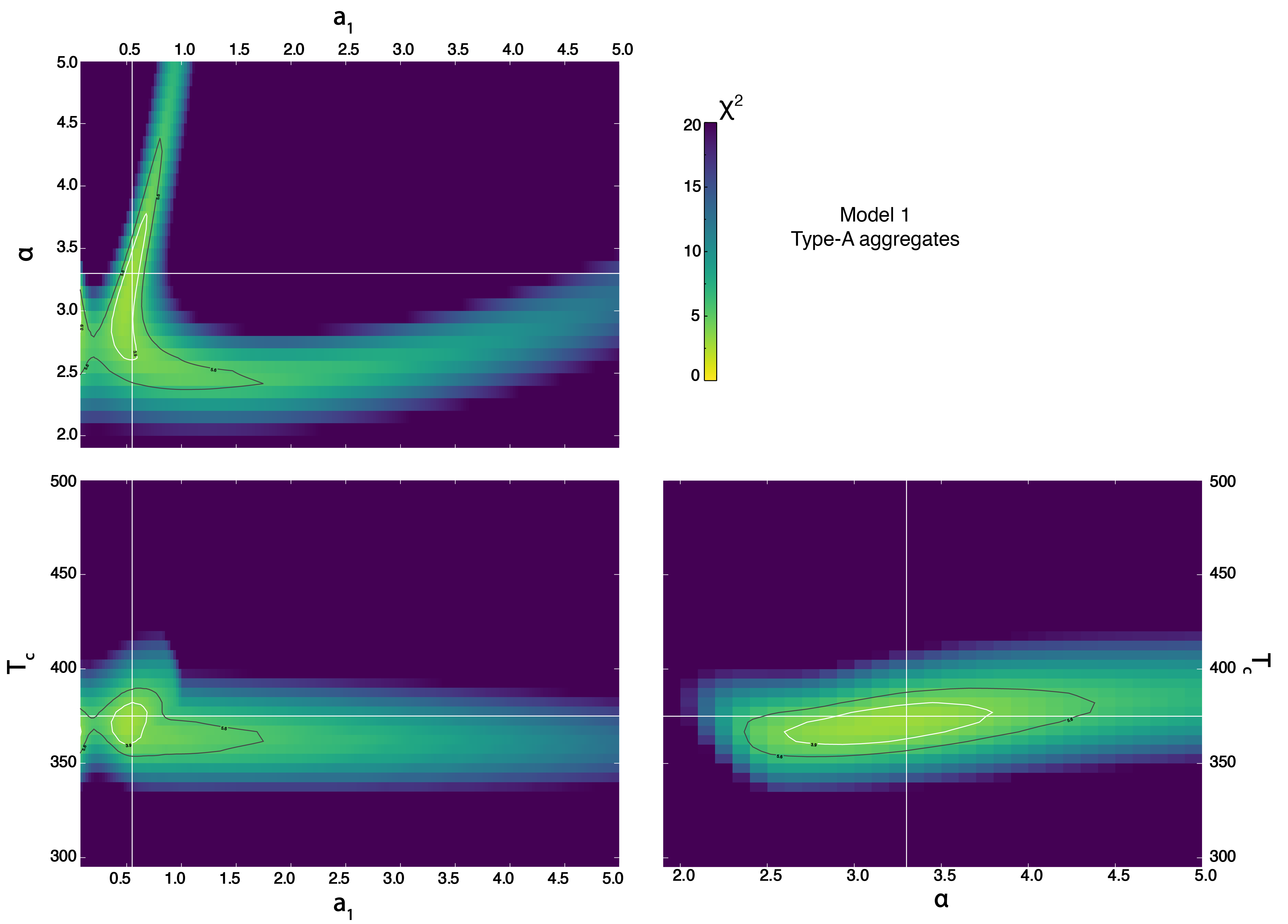}
  \caption{2D $\chi_{\nu}^{2}$ distributions resulting from the parameter exploration
    of the fit to the prism+LXD\_long spectrum in the case of an ice-free model (\textit{model 1}, Figure \ref{fig:prism_lxd_observations}). $T_{c}$, $a_{1}$, and $\alpha$ are the color temperature, the minimum radius in the particle size distribution, and the differential dust size distribution power-law index, respectively. In each panel, for each point in the grid, we search for the value of the third parameter the minimizes $\chi_{\nu}^{2}$, and the minimum $\chi_{\nu}^{2}$ is displayed.
    The solid curves shown in white and gray with
    $\Delta\chi_{\nu}^{2}$ = 1.00 and 2.71 represent the 68\% and 90\% confidence levels,
    respectively. Horizontal and vertical lines indicate the best-fit parameter values. The best fit to the data is found for $T_{c}$ = $375_{10}^{10}$~K, $a_{1}$ = $0.55_{0.15}^{0.15}$~$\mu$m, and $\alpha$ = $3.3_{0.6}^{0.5}$. The uncertainties on the best fit parameter values correspond to 1-$\sigma$ errors. \label{model1_joint_error_distribution}}
\end{figure*}
\vspace{1cm}
\subsection{Ice free model: Properties of the dust coma}\label{subsec:Ice free model: Properties of the dust coma}
Given the absence of any ice feature, we first model the spectrum by a power-law size distribution of porous dust aggregates only (type-A aggregates). This implies that $F_{A}$ is set equal to 1 in Eq.~\ref{eq:areal_mixture}.

There is a broad literature investigating the dust mineralogy of cometary comae and is based on several techniques, including but not limited to, laboratory studies of IDPs \citep{Wooden2000}, micrometeorites \citep{Bradley1999}, and STARDUST grains \citep{Brownlee2006};  analysis of ground and space-based mid-IR observations \citep{Harker2002, Woodward2021}; and in situ analysis of coma grains with spacecraft instrumentation \citep[e.g.,][]{Fomenkova1992, Fulle2016}. Given the featureless behaviour of the Wirtanen spectrum, the chemistry of the refractory component cannot be unambiguously identified. However, it is possible to derive information on the size distribution of the dust, once its chemical nature is assumed. We follow the same approach adopted by \citet{BM2017a,BM2017b} and consider amorphous carbon and amorphous olivine as coma refractories, with a silicate to carbon ratio of 0.3, based on the results by \citet{Fulle2016} at 67P/Churyumov-Gerasimenko (hereafter 67P). We adopt the optical constants by \citet{Edoh1983} and \citet{Dorschner1995} for amorphous carbon and amorphous olivine (Mg = Fe = 0.5), respectively.

The free parameters in the model are the exponent, $\alpha$, and the minimum radius, $a_{1}$, of the particle size distribution as well as the color temperature $T_{c}$. Because large particles do not contribute significantly to infrared radiation \citep[e.g.,][]{Harker2002}, the grain size distribution is truncated at 100~$\mu$m in diameter. This step reduces computation time. The fractal porosity $D$ is set to 2.5 \citep{BM2017a,BM2017b}. We acknowledge that this parameter was initially set free, and found it to be unconstrained by our data. We use a Levenberg–Marquardt $\chi^{2}$ minimization algorithm to find the best fit solutions \citep{Markwardt2009}. The model quality is quantified by means of the reduced $\chi_{\nu}^{2}$, defined as
\begin{equation}
  \chi_{\nu}^{2}=\frac{1}{\nu}\sum_{i=0}^{N}\Bigg(\frac{\mathrm{R}_{i}-\mathrm{R}_{i,model}(\alpha,a_{1},T_{c})}{\mathrm{ERR}_{i}}\Bigg)^{2},
\end{equation}
where $\mathrm{R}_{i}$, $\mathrm{ERR}_{i}$, and $\mathrm{R}_{i,model}$ represent the measured relative reflectance, the error on the relative reflectance, and the modeled relative reflectance, respectively, in correspondence of the individual wavelength, identified by the subscript $i$. The sum is over all $N$ wavelength points selected for the best fit minimization. The degree of freedom, $\nu$, equals the number of spectral elements $N$ minus the number of fitted parameters.

\begin{deluxetable*}{@{\extracolsep{8pt}}lclclclclccccc}
  \tablecaption{Properties of the refractory (type-A) and ice aggregates (type-B) in the coma of 46P/Wirtanen as derived from spectroscopic modeling. Model 1 corresponds to an ice free model and accounts for aggregates of amorphous carbon and amorphous olivine into vacuum (type-A aggregates). Model 2 is an areal mixture of type-A aggregates and US10-like water-ice grains (1.2-$\mu$m diameter monodisperse water-ice grains with a small, 0.5\%, dirt fraction by volume). Model 3 is an areal mixture of type-A aggregates and H2-like water-ice grains (1.0-$\mu$m diameter monodisperse pure water-ice grains). \label{table:modeling}}
  \tablewidth{0pc}
  \tabletypesize{\scriptsize}
  \tablehead
  {
    \colhead{}&
    \multicolumn{5}{c}{Type-A aggregates (refractory)}&
    \multicolumn{4}{c}{Type-B aggregates (icy)}&
    \colhead{$T_{c}$}&
    \colhead{$\chi_{\nu}^{2}$}\\
    \cline{2-6} \cline{7-10}
    \colhead{Model}&
    \colhead{$F[\%]$\tablenotemark{\scriptsize{a}}}& \colhead{$|\alpha|$\tablenotemark{\scriptsize{b}}} & \colhead{$D$\tablenotemark{\scriptsize{c}}} & \colhead{$a_{1}[\micron]$\tablenotemark{\scriptsize{d}}} & \colhead{$a_{2}[\micron]$\tablenotemark{\scriptsize{e}}} &
    \colhead{$F[\%]$}& \colhead{$|\alpha|$} & \colhead{$D$} & \colhead{$a [\micron]$} &
    \colhead{}&
    \colhead{}
  }
  \startdata
  Model 1&
  100\tablenotemark{\scriptsize{*}}& $3.3_{0.6}^{0.5}$ & $2.5$\tablenotemark{\scriptsize{*}}  & $0.55_{0.15}^{0.15}$ & $100$\tablenotemark{\scriptsize{*}}&
  0\tablenotemark{\scriptsize{*}} & ... & ... &  ...&
  $375_{10}^{10}$ &
  2.9\\[0.2cm]
  Model 2&
  $>97.8$ & 
  $3.3$\tablenotemark{\scriptsize{*}} & 
  $2.5$\tablenotemark{\scriptsize{*}} & 
  $0.55$\tablenotemark{\scriptsize{*}}& 
  $100$\tablenotemark{\scriptsize{*}} &
  $<2.2$ & 
  $0$\tablenotemark{\scriptsize{*}} & 
  $3$\tablenotemark{\scriptsize{*}} &  
  $0.6$\tablenotemark{\scriptsize{*}} &
  $375$\tablenotemark{\scriptsize{*}} &
  9.1\\
  Model 3&
  $>98.6$ & 
  $3.3$\tablenotemark{\scriptsize{*}} & 
  $2.5$\tablenotemark{\scriptsize{*}} & 
  $0.55$\tablenotemark{\scriptsize{*}}& 
  $100$\tablenotemark{\scriptsize{*}} &
  $<1.4$ & 
  $0$\tablenotemark{\scriptsize{*}} & 
  $3$\tablenotemark{\scriptsize{*}} &  
  $0.5$\tablenotemark{\scriptsize{*}} &
  $375$\tablenotemark{\scriptsize{*}} &
  9.3\\
  \enddata
  \tablenotetext{*}{This parameter has been set as constant.}
  \tablenotemark{\scriptsize{a}}{Fraction of the area occupied by the aggregates}
  \tablenotemark{\scriptsize{b}}{Differential size distribution power-law index}
  \tablenotemark{\scriptsize{c}}{Fractal porosity}
  \tablenotemark{\scriptsize{d}}{Minimum particle radius in the size distribution}
  \tablenotemark{\scriptsize{e}}{Maximum particle radius in the size distribution}
\end{deluxetable*}

To ensure that the solution obtained from the best-fit minimization algorithm is that of a global minimum rather than a local minimum and to investigate the possible inter-correlation between the free parameters, we explore a wide parameter space ($\alpha$,$a_{1}$), ($\alpha$,$T_{c}$) and ($a_{1}$,$T_{c}$) with $\alpha$ ranging between 2 and 5 with a step of 0.1, $T_{c}$ between 300~K and 500~K with a step of 5~K, and $a_{1}$ between 0.1 and 5~$\mu$m with a step of 0.025~$\mu$m. We determine the reduced $\chi_{\nu}^2$ distributions and the degree of confidence in the best-fit values. The joint error distribution of ($\alpha$,$a_{1}$), ($\alpha$,$T_{c}$) and ($a_{1}$,$T_{c}$) is shown in Fig.~\ref{model1_joint_error_distribution}. For each point in the grid, which defines the values of the two parameters under consideration, we solve for the third parameter. The 2D contours of constant $\chi_{\nu}^2$ corresponding to 68\% and 90\%  confidence level are displayed in white and gray, respectively. Best-fit parameter values and the correspondent 1-$\sigma$ errors are listed in Table \ref{table:modeling}. There is evidence for a slight correlation between the exponent of the particle size distribution, $\alpha$, and minimum size of the particle distribution, $a_{1}$, but both parameters are well constrained. To exclude the locations where emission bands occur and regions of strong telluric absorption, the best-fit optimization is performed in the wavelength ranges between 1.16--1.34~\micron, 1.45--1.79~\micron, 1.92--2.53~\micron, 2.94--3.20~\micron, 3.55--4.18~\micron, and 4.52--5.34~\micron.  The best-fit result (Figure \ref{fig:prism_lxd_observations}, red solid line, \textit{model 1} in Table \ref{table:modeling}) has a differential dust size distribution power-law index of 3.3, minimum size $a_{1}$ equal to 0.55~$\mu$m and a color temperature of 375~K. The derived color temperature is in excess with respect to the equilibrium temperature $T_{eq} = 278 \rh^{-0.5} =271~K$. This could be attributed to the presence of $\mu$m- and sub-$\mu$m sized grains comprising absorbing material. However, we report in Section \ref{subsec:nucleus} that the thermal emission has a substantial contribution from the nucleus, which is challenging to remove. Therefore, we do not interpret this result any further.

\subsection{An upper limit for water ice}\label{subsec:Upper limit for the water-ice abundance}
\begin{figure}[htp]
  \includegraphics[width=\linewidth]{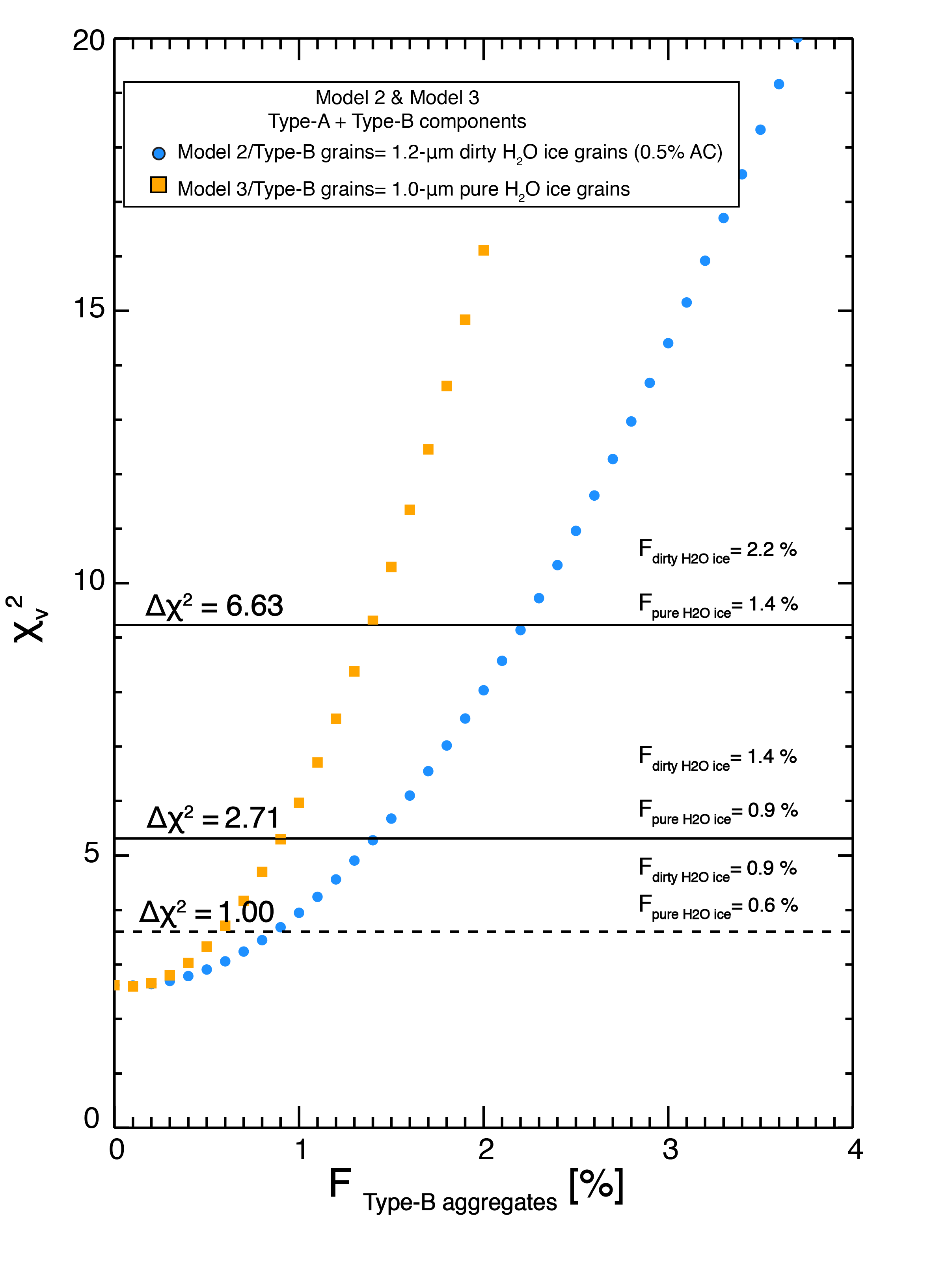}
  \caption{Upper limits for the abundance in the coma of Wirtanen of 1.2-$\mu$m diameter water-ice grains with 0.5\% amorphous carbon by volume (spectral model 2, blue dots) and 1.0-$\mu$m diameter pure water-ice grains (spectral model 3, orange squares). The upper limit for the water-ice fraction $F$ is given at the 68\%, 90\%, and 99\% confidence levels for both models. Models 2 and 3 differ exclusively for the physical properties (purity and particle size) of the ice grains (type-B grains), as the properties of the refractory aggregates (type-A particles) are the same in both simulations (see text for details). \label{water_ice_upper_limit}}
\end{figure}

Our data do not display spectral evidence of water ice in the coma. Nevertheless, this does not necessarily rule out the presence of water ice in the coma of Wirtanen. We investigate this possibility and assess an upper limit for water-ice abundance by modeling the composite spectrum prism+LXD\_long acquired on 2018 December 13 (Figure \ref{fig:prism_lxd_observations}) with two components (see Sec. \ref{subsec:Modeling}), one of polydisperse porous dust aggregates (type-A aggregates)  and one of polydisperse water-ice particles (type-B aggregates).  We hold fixed the physical properties of the dust aggregates using the best-fit values from the ice-free model (\textit{model 1}, Section \ref{subsec:Ice free model: Properties of the dust coma}). This limits the number of free parameters. Additionally, given the lack of water-ice absorption features and therefore the impossibility to solve for both the physical properties of the ice grains and their abundance, we assume that water-ice particles with the same properties as those found in comet C/2013 US$_{10}$ (model 2) and 103P/Hartley 2 (model 3) were ejected by comet Wirtanen. Therefore, either 1.2~\micron~in diameter ice grains containing 0.5\% amorphous carbon by volume \citep[model 2,][]{Protopapa2018} or 1.0~\micron~in diameter pure ice grains \citep[model 3,][]{Protopapa2014} are assumed to be present in the coma of Wirtanen. We consider the type-A and type-B components separated from one another (see Eq. \ref{eq:areal_mixture}). The free parameter in the model is the fraction of type-B grains in \%. We varied the fraction of type-B grains ($F_{B}$, or equivalently $1-F_{A}$) between 0 and 1 with a step interval of 0.1\%. For each value of $F_{B}$ we computed $\chi_{\nu}^{2}$ as discussed in Section \ref{subsec:Ice free model: Properties of the dust coma}. In Figure \ref{water_ice_upper_limit} we present a plot of $\chi_{\nu}^{2}$ versus $F_{B}$. The two curves corresponding to model 2 (blue dots) and model 3  (orange squares) have a minimum for $F_{B}=0.1\%$, which is almost identical to \textit{model 1} (red line in Figure \ref{fig:prism_lxd_observations}). However, in the case of US10-like water-ice grains (model 2), the data are consistent with an upper limit of water-ice abundance on the order of 1.4\% and 2.2\% at the 90\% and 99\% confidence levels, respectively \citep[for 1 free parameter, $\Delta \chi^{2}=2.71$ and $6.63$ corresponds to a confidence level of 90\% and 99\%, respectively,][]{Avni1976}. In the case of pure water-ice grains (H2-like water-ice grains, orange squares in Figure \ref{water_ice_upper_limit}) an upper limit of 0.9\% and 1.4\% at the 90\% and 99\% confidence levels, respectively, is estimated. Model 2, with an upper limit of 2.2\% for the water-ice abundance is shown in Figure~\ref{fig:prism_lxd_observations} (gray line). Model 3 with a 1.4\% water-ice fraction is almost identical to model 2 and therefore it is not shown in Figure~\ref{fig:prism_lxd_observations}. The upper limit of water-ice abundance is systematically lower in model 3 with respect to model 2. This is because pure water ice displays stronger water-ice absorption bands with respect to ice grains with refractory inclusions. We point out that the band depth of water-ice absorption features increases with increasing path length or equivalently particle diameter. Therefore, assuming water-ice particles larger than those observed in the coma of US10 and H2 would lead to even lower upper limits of the water-ice areal fraction. The upper limit for the fractional area of US10-like and H2-like water-ice grains of 2.2\% and 1.4\% translates into an ice-to-dust areal ratio of 3\% and 2\%, respectively. These values are estimated accounting for the 15\% nucleus contribution to the scattered component of the spectrum (Section \ref{subsec:nucleus}). We conclude that the coma of Wirtanen is dominated by dust. If water-ice grains were ejected by comet Wirtanen, their abundance would be negligible.
\subsection{Water-ice grain sublimation lifetime}\label{subsec:Water-ice grain sublimation lifetime}

In order for water-ice grains to be a candidate for hyperactivity, they must be ejected in high enough abundance and sublimate quickly enough to account for the observed total water production rates.  We did not detect any water-ice features in our spectra, therefore we additionally require the grains to fully sublimate before they travel $\sim$1\arcsec{} from the nucleus, this distance based on our spectral aperture sizes.

\begin{figure}[htp]
  \includegraphics[width=\linewidth]{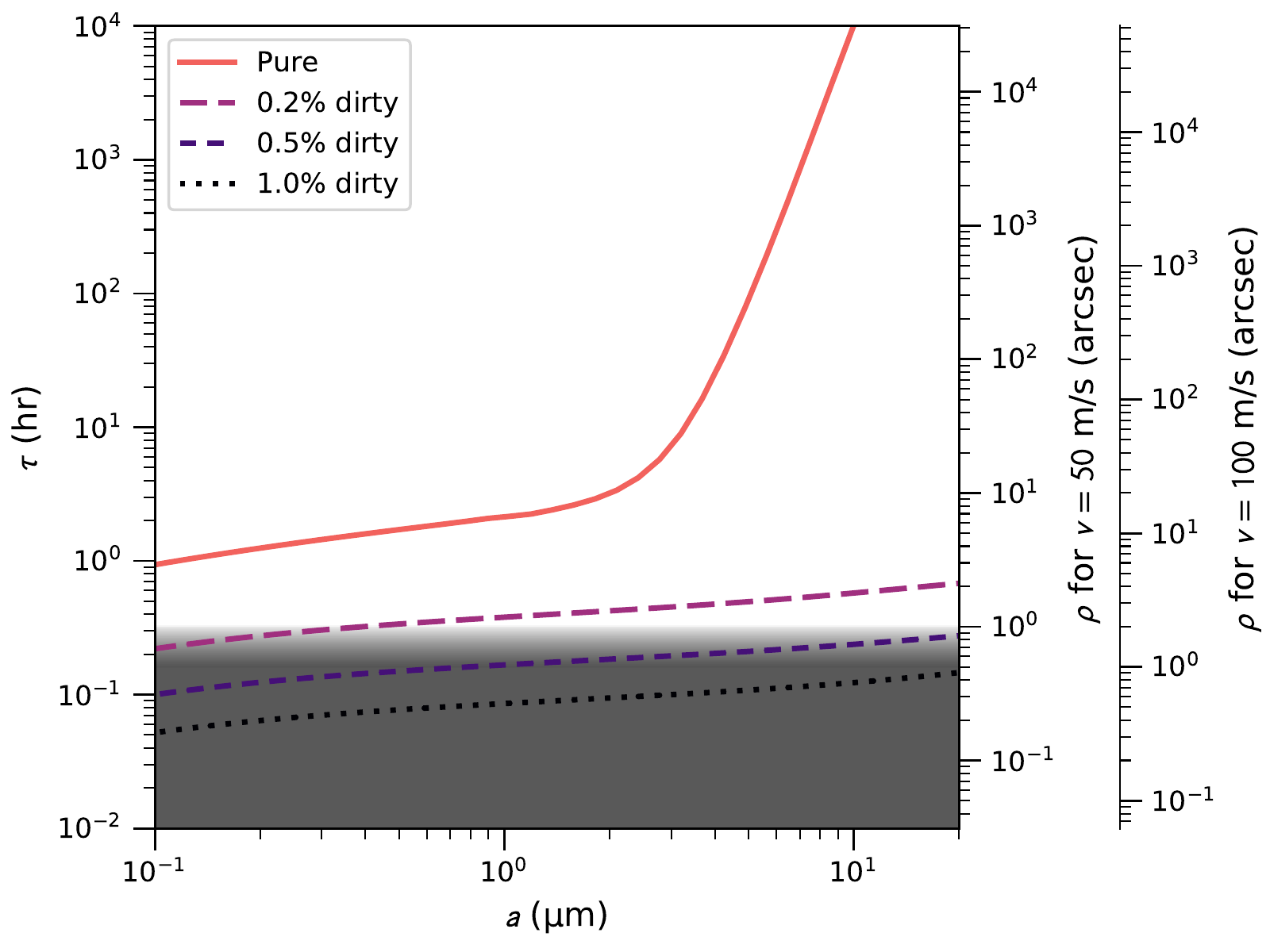}
  \caption{Water-ice grain lifetimes at 1.055~au for pure water ice, and mixtures of water ice with 0.2, 0.5, and 1.0\% amorphous carbon dust by volume.  The right axes indicates the distance in arcseconds at $\Delta= 0.08$~au traveled before sublimating for two coma expansion speeds: 50 and 100~m~s$^{-1}$, based on grain speeds estimated by \citet{Fulle2000}.  Model lines which fall in the shaded regions are more likely to produce a dust dominated spectrum, as we observed at 46P. \label{fig:lifetime}}
\end{figure}

Pure ice grains have the longest sublimation lifetimes, and dust impurities tend to reduce that lifetime.  \citet{Protopapa2018} modeled grain lifetimes following \citet{Hanner1981}, \citet{Lien1990}, and \citet{Beer2006}, and we adopt this model for the analysis of comet Wirtanen.  It computes grain temperatures based on the absorption of sunlight, grain thermal emission, and cooling from sublimation.  Radiative transfer calculations are based on Mie theory \citep[code of][modified by B.~Draine]{Bohren1983}\footnote{Available at \url{https://www.astro.princeton.edu/~draine/scattering.html}}.  Mass loss by sublimation decreases the grain radius, as does sputtering from the solar wind.  Grain radii are integrated down to 10~nm, at which point they are considered destroyed.  We present calculations for mixtures of water ice \citep[optical constants by][]{Warren2008} and amorphous carbon \citep{Edoh1983}, using effective medium theory \citep[Bruggeman mixing rule;][]{Bohren1983}.  Calculations for water-ice grain lifetimes at 1.055~au with 0, 0.2, 0.5, and 1\% amorphous carbon by volume as a function of grain diameter are presented in Fig.~\ref{fig:lifetime}, with an initial grain diameter of 20~\micron.

Based on our calculated water-ice grain lifetimes, micrometer-sized pure water-ice grains have a lifetime of 1 to 3~hr at 1.055~au, and dirty ice grains have lifetimes that are shorter by an order of magnitude or more.  Beyond 3~\micron{} in size, the difference is even more dramatic.  Pure-ice grain lifetimes increase to $10^4$~hr for 10-\micron{} diameter grains, whereas dirty-ice grain lifetimes are still less than 1~hr, even for the low carbon fraction of 0.2\%.

To further distinguish between the type of ice (pure versus dirty) that could be present in the inner coma of comet Wirtanen, we need to consider the time it takes for a coma grain to travel to the edge of the spectral aperture in projection. Faster coma expansion speeds require rapid sublimation rates in order to account for the lack of water-ice absorption features in our spectra. \citet{Fulle2000} estimated several dust parameters, including expansion speed, for comet Wirtanen with a dynamical model of dust and an image of the comet near its 1997 perihelion. They estimated expansion speeds for 1~\micron{} diameter dust to be 65 to 95~m~s$^{-1}$, and we assume the dust and ice grain expansion speeds are the same. The aperture crossing times for these speeds are 0.17 to 0.25~hr for $\Delta= 0.08$~au. In Fig.~\ref{fig:lifetime}, we provide alternative y-axes showing the distance traveled by a grain moving at 50 and 100~m~s$^{-1}$ in the image plane.

Water-ice grains with sublimation lifetimes longer than the aperture crossing timescales (i.e., $\gtrsim$1~hr) would have been detected in our spectra if their abundance was higher than a few percent (Section~\ref{subsec:Upper limit for the water-ice abundance}).  Assuming that higher abundances are needed to account for the hyperactivity, we rule out pure water ice grains, which have lifetimes $\gtrsim1$~hr for diameters $\gtrsim0.1$~\micron{}, and therefore would survive the time of flight to the edges of our spectral apertures.
In contrast, dirty-ice grains are likely to have sublimated before reaching the edge of our LXD spectral aperture.  Given that \citet{Fulle2000} estimated expansion speeds from 65 to 95~m~s$^{-1}$ for 1~\micron{} grains, we consider dirty ice with carbon fractions $\gtrsim0.5$\% as candidates for explaining the hyperactivity of comet Wirtanen.
\subsection{Coma reddening}\label{subsec:Coma reddening}
\begin{figure*}[htp]
  \includegraphics[width=\linewidth]{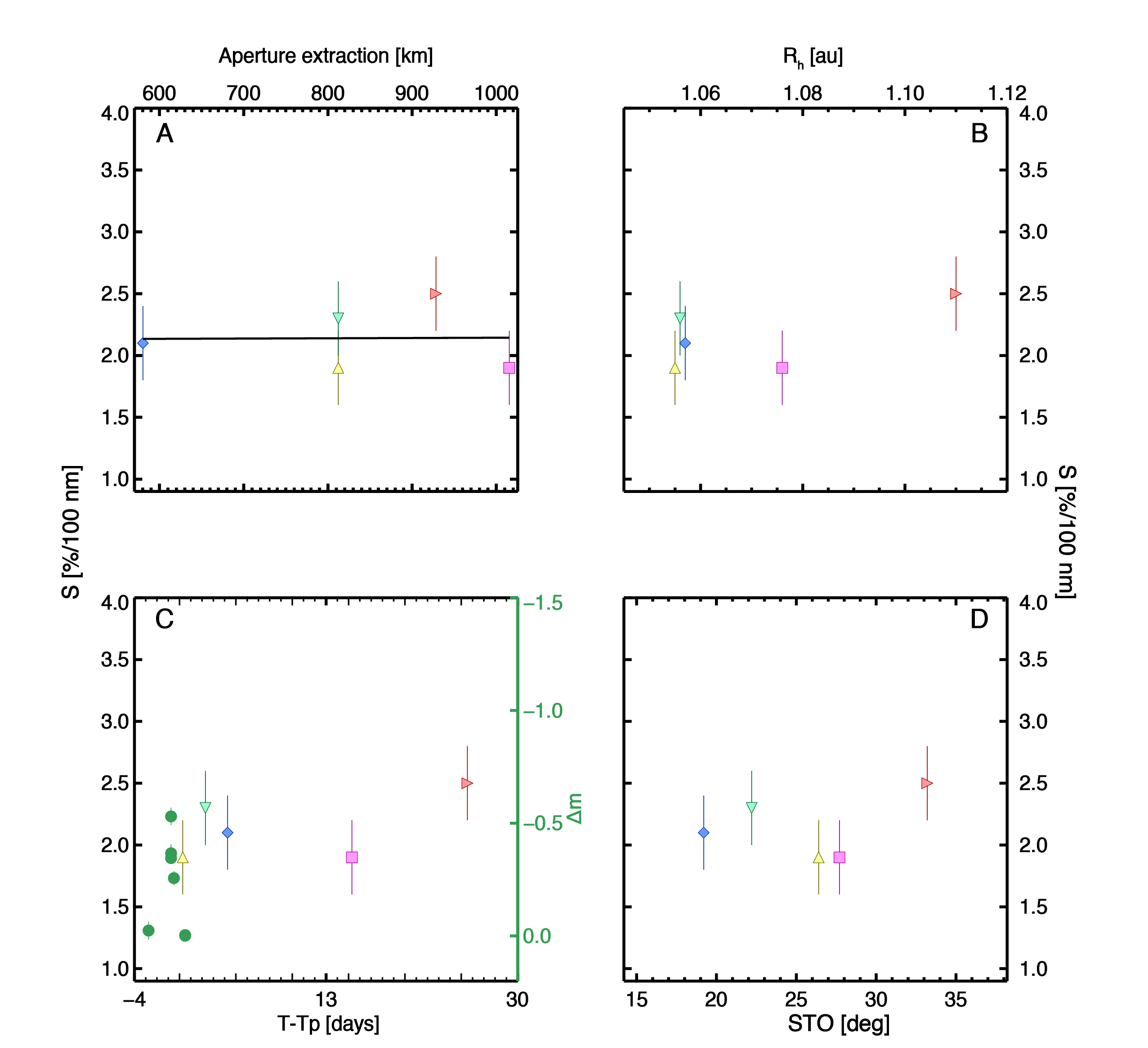}
  \caption{Dust coma spectral gradient of comet 46P/Wirtanen expressed in \%/100~nm as a function of projected aperture diameter in km (panel A), heliocentric distance $R_{h}$ (panel B), observing time since perihelion (panel C) and Sun-target-observer (STO) angle (panel D). The comparison between the spectral gradient behavior as a function of time and comet Wirtanen light curve (green dots) during an outburst close to perihelion discussed in detail by \citet{Kelley2021} is also shown in panel C. \label{fig:coma_reddening}}
\end{figure*}

The color of the dust coma of comet Wirtanen ($S$, expressed in \%/100~nm) as inferred from the IRTF SpeX measurements (Section \ref{sec:Observations}) in the near-infrared wavelength range lacks variation across our observational data set (Table \ref{tab_obs}). To investigate its significance, we considered putative correlations between $S$ and aperture extraction, heliocentric distance, time, and phase angle (Figure \ref{fig:coma_reddening}).

First we looked at whether the spectral color might be influenced by the spectral aperture in projection $d$ (in km, $d\sim \delta \Delta/206265$, where $\delta$  is the spectral aperture diameter in arcsec and $\Delta$ is the geocentric distance expressed in km). No color gradient is observed with aperture extraction (Figure \ref{fig:coma_reddening}, panel A). The linear fit to the data, shown in panel A as a solid line, has a $\chi^2_{\nu}$ = 1.0 and a slope of $10^{-5}$\% per 100~nm per km.

Color variations in cometary comae can be diagnostic of the coma dust properties, but also the ice-to-dust ratio \citep[e.g.,][]{Filacchione2020}. Water ice is stable at large heliocentric distances, but sublimation rates vary inversely with heliocentric distance. Our data (Figure \ref{fig:coma_reddening}, panel B) do not display any color variation with heliocentric distance. This is not unexpected given that the range of heliocentric distances covered by our observations is not wide enough to be sensitive to changes in ice-to-dust ratio.

Occasional variations in the spectral gradient could also be due to rapid changes in the coma composition and particularly in the ice-to-dust ratio as the result of impulsive events, such as outbursts, which eject fresh new material into the coma. The coma color is constant as a function of time from perihelion (Figure \ref{fig:coma_reddening}, panel C). An outburst was identified by \citet{Kelley2021} during the time period covered by our data with an apparent strength on the order of $-0.5$ mag (see green data points in panel C of Figure \ref{fig:coma_reddening}). If water-ice would be ejected during the outburst, we would expect a sudden blueing of the coma followed by a reddening due to sublimation.  However, the outburst ejecta moved rapidly, with speeds $>$23~m~s$^{-1}$, leaving little to no material near the nucleus at the time our spectrum was taken \citep{Kelley2021}.  In addition, anticipating the post-ejection evolution in the ice properties is not straightforward.  But no significant trend is observed in the color gradient after the time of the outburst.

Finally, no correlation between the coma color and phase angle (Sun-Target-Observer angle, STO in Table~\ref{tab_obs}) over the range between 20$^{\circ}$ and 50$^{\circ}$ is observed (Figure \ref{fig:coma_reddening}, panel D). An increase of reddening with phase angle for cometary dust has been reported for the first time by \citet{BM19} for comet 67P. Specifically, \citet{BM19} reported a phase reddening in the near-IR of the dust coma in the near-nucleus of 67P of 0.031\%/100~nm~deg$^{-1}$ in the phase angle range 50$^{\circ}$-120$^{\circ}$. This phenomena has been attributed by \citet{BM19} to the roughness of the dust particles \citep{Beck2012,Schroder2014}. As observed by \citet{Hartmann1984}, changes in the water-ice-to-dust ratio could produce an apparent phase reddening due to the relationship between phase angle and heliocentric (or observer-target) distance. A comet observed at large heliocentric distances and therefore low phase angles would display a coma color that is bluer with respect to the same comet observed closer-in at larger phase due to the sublimation lifetime of the ice grains. Therefore, to test for the presence of a true phase reddening and rule out the hypothesis of compositional (ice-to-dust ratio) changes, the correlation between spectral gradient and heliocentric distance needs to be tested as well. \citet{BM19} reported no significant heliocentric variations
in the dust color of 67P, which reinforces the result of a true phase reddening. No phase reddening is observed in our data.

\vspace{1cm}
\section{Summary and discussion}\label{sec:Discussion and conclusions}
We report a detailed characterization of the inner coma of comet 46P/Wirtanen through infrared spectroscopic measurements acquired with IRTF SpeX during the close approach of the comet to the Earth in December 2018. Our findings are:
\begin{enumerate}
  \item{No water-ice absorption features are detected in our data, which span a range of heliocentric distance from 1.1~au inbound, to 1.05~au near perihelion, and back to 1.1~au outbound.}

  \item{The modeling analysis of the observations acquired immediately post perihelion and covering the wavelength range 0.8--5.3~\micron{} indicates the coma is composed of porous refractory aggregates (fractal porosity assumed equal to 2.5) spanning 1.1--100~\micron{} in diameter with a differential dust size distribution power-law index of --3.3$\pm$0.5 and a color temperature of 375$\pm$10~K.}
  
  \item{An upper limit of 1.4\% in water-ice abundance is estimated, assuming water-ice grains with physical properties identical to those observed in the coma of the hyperactive comet H2 \citep[1.0-$\mu$m in diameter pure water-ice grains][]{Protopapa2014}. A water-ice fraction up to 2.2\% is consistent with Wirtanen data in the case of water-ice grains of 1.2~\micron{} in diameter with 0.5\% amorphous carbon by volume, like those observed in the coma of US10, a possible but unconfirmed hyperactive comet \citep{Protopapa2018}. Lower carbon fractions and/or larger water-ice particle diameters would lower this upper limit.}
  
  \item{No significant variations in spectral slope are observed among our spectral data.}
  
\end{enumerate}

Comet Wirtanen was identified as a hyperactive comet by \citet{Lamy1998-Wirtanen}, \citet{GroussinLamy2003} and \citet{Lis2019} with an active fraction between 60\% and 120\% at perihelion during the 1997 and 2002 apparitions. However, water production rate can vary from apparition to apparition, as was the case for comet H2 \citep{Combi2011}. Therefore, to put into context our observations and test whether hyperactivity is associated with the presence of a water-ice grain halo, it is critical to first estimate the active fraction of comet Wirtanen during the 2018 apparition. For this purpose, we used the sublimation model by \cite{Cowan1979} for a rotational pole pointed at the Sun\footnote{Code available at \url{https://github.com/Small-Bodies-Node/ice-sublimation}}.  This is the same model as employed by \citet{Lis2019}. Adopting the mean nucleus radius of 555$\pm$40~m estimated by \citet{Boehnhardt2002}, we translated the water production rate of $Q_\mathrm{{H_{2}O}}=7.2\times10^{27}$ molecules s$^{-1}$ reported by \citet{Knight2021} on 2018 December 16 ($\mathrm{Z} = 3.19\times10^{21}~ \mathrm{molecules~s^{-1}~m^{-2}}$ at $\mathrm{R_{h}}$=1.056~au), close in time to our spectroscopic measurements, into an active fraction of 58\%. The water production rate reported by \citet{Knight2021} and derived from measurements of the OH 309~nm (0-0) band is in agreement with the H$_{2}^{16}$O production rate of (7.7$\pm$1.5)$\times$10$^{27}$ molecules s$^{-1}$ quoted by \citet{Lis2019} and based on observations of the $1_{1,0}-1_{0,1}$ transition of H$_{2}^{18}$O between 2018 December 14 and 20 UT. The water production rate was 10\% higher on 2018 December 3 \citep[8$\times$10$^{27}$ molecules s$^{-1}$;][]{Knight2021}, which translates into an active fraction of 65\%. This active fraction is similar to that of comet H2 during the Deep Impact observations \citep[$\sim$72\%,][]{Lis2019} and it is at least 2 times higher than the typical active fraction for most comets \citep[e.g.,][]{AHearn1995}.

In this paper, we test the hypothesis put forth by \citet{AHearn2011}, \citet{Protopapa2014}, and \citet{Kelley2013} that sublimating water ice in the coma in the form of small grains and/or large chunks is responsible for a comet hyperactivity. We point out that \citet{Knight2021} and \citet{Bonev2021} both provide indirect evidence for the presence of an extended source of water vapor in the coma of Wirtanen. However, no information on the properties of the ice that produces the water vapor, such as particle size and ice-to-dust ratio, has been retrieved based on their work.

Radiative transfer modeling of the spectroscopic data combined with calculations of water-ice grain sublimation lifetimes rule out H2-like small water-ice grains \citep{Protopapa2014} in the coma of Wirtanen. The long lifetime of 1-$\mu$m in diameter pure water-ice particles is inconsistent with an upper limit of only 1.4\% for the abundance of these grains in the coma and the assumption that high water-ice mass-loss is needed to account for the hyperactivity. We consider instead dirty small ice grains (1.2-$\mu$m in diameter) with carbon fractions $\gtrsim$0.5\% as possible candidates to account for the comet water production rates. The small amount of low albedo dust limits the observability of the water-ice grains at heliocentric distances of 1.0–1.1~au, accounting for our observations. Moreover, it quickly converts the ice into water vapor, which is needed to account for the hyperactivity. This model is based, in part, on the ice grains observed in the coma of US10, which has been suggested to be an hyperactive comet by \citet{Protopapa2018} based on the detection of water-ice grains in the coma presumably ejected by sublimation of the hyper-volatile CO$_{2}$, a process observed at H2 \citep{AHearn2011,Protopapa2014}. Interestingly, CO$_{2}$ has been detected in comet Wirtanen as well \citep{Bauer2021}. Notice that the detection of CO$_{2}$ does not necessarily imply a water-ice grain halo. As an example, while CO$_{2}$ has been mapped in the coma of comet 67P \citep{Fink2016}, a regular comet in terms of level of activity (non hyperactive), no water-ice grains have been detected in the quiescent coma of the comet \citep{BM2017a,BM2017b}. Small water-ice grains on the order of a $\mu$m (5 $\mu$m in diameter) have been reported by \citet{Davies1997} in the quiescent coma of comet Hale-Bopp, which is on the high end of a typical comet \cite[report an active fraction on the order of 34\% at 0.91~au]{Lis2019}. We stress that the water-ice physical properties (purity, grain size, abundance) depend on the radiative transfer model applied to the spectroscopic measurements. Therefore, a consistent comparison between H2, US10, Wirtanen and Hale-Bopp requires the same modeling strategy to be applied to all the data. We reserve this investigation for future work.

\citet{Kelley2013} examined a population of point sources surrounding the nucleus of comet H2 in Deep Impact images.  \citet{Kelley2015} revised their estimates and found sizes up to 8~m in diameter for the point sources and suggested they could be responsible for the comet's extended water vapor production and hyperactivity.  Radar observations are sensitive to particles of the same size range.  In fact, the radar cross section of $\gtrsim2$~cm particles is the same order of magnitude as the one reported by \citet{Kelley2015} for the point sources,  assuming nucleus-like properties (e.g., low albedo). This suggests that the two populations, those in the Deep Impact images and those in the radar, could be related.  Radar observations of comet Wirtanen in 2018 also indicate the presence of centimeter-sized grains in the coma.  The total cross section of grains with radii $\gtrsim2$~cm is roughly the same than that observed at comet H2, 0.89~km$^2$,  \citep[][and E.\ Howell, private communication]{Harmon2011}.

If the centimeter-sized radar observed particles would be responsible for the hyperactivity of comet Wirtanen, then the lack of water-ice spectral signatures in our observations suggests that the centimeter-sized grains are either a small areal fraction of the coma or are not icy on the surface.  Putting the Wirtanen and H2 data together, the one possible scenario that can explain all observations and account for the hyperactivity of both comets requires large centimeter-to-decimeter sized particles that behave like mini-cometary nuclei, i.e., the water-releasing materials would reside just under the surface at a depth deeper than that sampled by near-IR remote sensing (i.e., on the order of $\mu$m). However, this does not account for any contribution from the fine-grained water ice observed near the nucleus of H2 \citep{AHearn2011,Protopapa2014}. We point out that large chunks were observed in the coma of 67P by \citet{Rotundi2015}, \citet{Fulle2015ApJ,Fulle2016ApJ} and \citet{Agarwal2016}, nevertheless this comet is not hyperactive.

In summary, we provide observational limits on the physical properties of the icy grains responsible for hyperactivity in comet Wirtanen: either icy grains on the order of 1~\micron{} in size with a small amount of dust, or large chunks containing significant amounts of water ice.


\acknowledgements
The authors thank D.~Wooden for helping to identify the cometary CN emission.  Support for this work was provided through the NASA Solar System Observations program grant number 80NSSC20K0673. S.P. thanks the NASA grant 80NSSC19K0402 and the Space Telescope Science Institute grant (HST-GO-15372) for partial funding that supported her work. C.E.W. acknowledges partial support from NASA Solar System Observations grant 80NSSC19K0868. We thank the anonymous referee and D. Bockel\'{e}e-Morvan for valuable comments that improved the manuscript.

\facilities{NASA IRTF (SpeX)}

\software{astropy \citep{Astropy2018}, JPL Horizons \citep{Giorgini1996}, IDL, Spextool \citep{Cushing2004} }

\bibliography{Protopapa_46P}{}
\bibliographystyle{aasjournal}



\end{document}